\documentclass[12pt]{article}

\usepackage{graphicx}
\usepackage{color}
 
\begin{document}
 \title{Effect of a Pseudogap in Strong Coupling d-Wave Superconductivity.}
 \author{L. Coffey,\\
 Physics Department,\\ 
 Illinois Institute of Technology,\\
 Chicago, Illinois 60616}
\maketitle

\begin{center}
{\bf Abstract}
\end{center}

A phenomenological strong coupling model that has been used to analyze
superconducting-insulator-superconducting (SIS) break junction experiments
on optimal to overdoped Bi$_{2}$Sr$_{2}$CaCu$_{2}$O$_{8}$ (Bi-2212) is modified to include a pseudogap. The calculated density of states and SIS conductances
are compared with experimental data on the underdoped phase of Bi-2212.

\newpage
The electronic properties of the copper oxides display evidence for two energy gaps: the superconducting gap, and the
pseudogap. The latter leads to a reduction in
the normal state density of states about the Fermi energy, and manifests itself in various normal
state thermal and transport properties \cite{Tim}. In addition, both normal and superconducting properties of the copper oxides
depend on carrier doping. The superconducting transition 
temperature $T_{C}$ varies strongly
with hole doping, increasing from zero in the overdoped phase, passing through a maximum at optimal doping, and
decreasing to zero in the underdoped phase. The curve describing this doping dependence is usually called the
superconducting dome. The temperature, usually denoted by  $T^{*}$, below which the pseudogap is non zero, also depends on doping.
The interplay between the pseudogap and the superconducting gap is currently an active area of study with 
several ideas being pursued to characterize the nature of the combined pseudogap/superconducting phase.
\cite{Huf,Tmr}\\   
\\
Point contact tunneling (PCT) and scanning 
tunneling microscopy (STM) experiments provide evidence for both gaps in the
electronic spectrum of the copper oxide superconductors. The hole doped material Bi$_{2}$Sr$_{2}$CaCu$_{2}$O$_{8}$ (Bi-2212) has been studied extensively with these techniques, and this superconductor will be
referred to in the present work.\\
\\
In PCT\cite{JZ}, the energy of the peaks in the conductance (dI/dV) curves of superconducting-insulator-superconducting (SIS) break junctions in Bi-2212 crystals yield superconducting gap values 
which correlate with the doping dependence of the superconducting transition temperature T$_{C}$ for optimal to overdoped samples. The measured superconducting gaps can also be understood semi-quantitatively using a
strong coupling superconductivity model \cite{LC}. However, in 
underdoped Bi-2212 crystals, the gap values measured from the SIS conductance dI/dV peaks continue to increase
to large values even as the T$_{C}$ values decrease to zero. This suggests that the
superconducting gap is not solely determining the dI/dV peak positions, but that the pseudogap is being
measured in this doping range. Furthermore because the measured pseudogap increases while at the same time T$_{C}$
is decreasing to zero, one interpetation of the underdoped SIS measurements is that the pseudogap is not a precursor superconducting pairing phenomenom,
but is instead a competing effect.\\
\\ 
In STM experiments \cite{Psh,Boy,Fuj}, the measured conductance (dI/dV) is closely related to the superconducting density
of states. In the optimal to overdoped Bi-2212 samples, the STM measurements reveal a conductance with a single peak, from which the value of the superconducting gap can be estimated. 
In the underdoped regime, two features are seen in the dI/dV curves: a low
energy shoulder-like feature which is likely associated with the superconducting gap, and
a high energy peak associated with the pseudogap, similar to the
SIS measurements.\\
\\
In the SIS measurements, the dip above the main dI/dV peak, which has been modelled as evidence of the superconducting pairing mechanism \cite{LC}, disappears in increasingly underdoped samples\cite{JZ}. This
behavior is also seen in some STM experiments\cite{Psh}.\\  
\\
Angle Resolved Photoemission Spectroscopy (ARPES) has provided 
information on the properties of both the superconducting gap and pseudogap along the Bi-2212 Fermi surface \cite{Yosh,Kondo}. One interpetation of these ARPES measurements leads to a pseudogap which has a maximum at the Fermi surface antinodal
point (the location of the maximum of the superconducting d-wave gap), and which decreases to
zero at a point between the antinodal point and the d-wave
superconducting gap node. In the normal state, this leads to a pseudogap
state involving a gapless
arc of states along the Fermi surface, centered about the nodal point of the d-wave
superconducting gap (which develops below T$_{C}$). The resulting normal state density of
states displays a reduction in states around the Fermi energy similar to experiment.
Another interpetation of ARPES measurements suggests the existence of pockets on the Fermi surface
in the pseudogap (underdoped) state \cite{Pdj}. This supports 
a theoretical model\cite{Yrz} based on the RVB theory which explains the pseudogap regime as
arising from a complex reconstruction of the Fermi surface involving 
the development of electron and hole pockets at the antinodal and nodal
regions respectively. The shrinking of the hole pockets with underdoping causes increasing pseudogap behavior in transport and thermal properties\cite{Carb1}. The interpetation of the ARPES and STM experiments
is an area of active research\cite{Koh,Pdc}.\\
\\  
Numerical results are presented here for the density of states, and the SIS conductance, in which a pseudogap of
non-superconducting origin is included in the strong coupling 
model that has been used to 
describe the optimal to overdoped regime of Bi-2212\cite{LC}. Extending this  model into
the underdoped phase of Bi-2212 is achieved via the magnitude chosen for the 
pseudogap in the calculation, assuming it to be directly determined by the doping level, and to increase
from zero at optimal doping as the superconductor becomes increasingly underdoped. 
An analogous connection between doping and the pseudogap magnitude is also
incorporated in other work\cite{Yrz}.  
In that approach, the pseudogap has a doping dependence
given by $0.3t(1-x/0.2)$ where $t$ is the tight binding hopping parameter, $x$ is
the hole doping, and $x=0.2$ is used to denote optimal doping from which the pseudogap
is assumed to start from zero (see figure 2(e) \cite{Yrz}).  \\
\\
The phenomenological strong coupling model used in this work captures many
experimentally measured features of tunneling data in the superconducting state of Bi-2212 over the whole doping range 
from under to over doped. The success of the present approach depends
on the form of the pairing spectral function, or {\em pairing glue}, used in the calculations. This is discussed in the following section of this manuscript.\\
\\
The equations are solved self consistently for the resulting 
superconducting gap and density of states. An increasing pseudogap
suppresses the superconducting gap to zero, and leads to features in the 
density of states and trends in the SIS conductance curves that are 
similar to the STM and PCT experiments.\\
\newpage
{\bf Theoretical Formalism}\\
\\
The strong coupling model equations to be solved self-consistently are given by\\
\begin{eqnarray}
\Delta(\omega) &=& \frac{1}{Z(\omega)}\displaystyle\int_{0}^{\omega_{c}}d\nu 
\int_{0}^{2 \pi} \frac{d \phi}{2 \pi} c_{\Delta} {\rm Re} \bigg\{  \frac{\Delta(\nu) {\rm cos}^{2}(2 \phi)}{[ \nu^{2} - \Delta^{2}(\nu) {\rm cos}^{2}(2 \phi) - \Delta_{PG}^{2}(\phi)]^{1/2}}\bigg\} \nonumber \\
&\times&
\displaystyle\int_{0}^{\omega_{c}}d\omega ' F(\omega ')   
\bigg[ \frac{1}{\omega + \omega ' + \nu + i\delta} - \frac{1}{\omega - \omega ' - \nu +i\delta}\bigg] \nonumber \\
\end{eqnarray}
and\\
\\
\begin{eqnarray}
[1-Z(\omega)]\omega &=& \displaystyle\int_{0}^{\infty}d\nu \int_{0}^{2 \pi} \frac{d \phi}{2 \pi} 
c_{Z} {\rm Re} \bigg \{ \frac{\nu}{[ \nu^{2} - \Delta^{2}(\nu){\rm cos}^{2}(2 \phi) - \Delta_{PG}^{2}(\phi) ]^{1/2}} \bigg \} \nonumber \\
&\times&
\displaystyle\int_{0}^{\omega_{max}} d\omega ' F(\omega ')
 \bigg [ \frac{1}{\omega + \omega ' + \nu + i\delta}
  + \frac{1}{\omega - \omega ' -\nu + i\delta}\bigg ] \nonumber \\
 \nonumber \\
\end{eqnarray}

Equations (1) and (2) are derived from the standard Eliashberg type equations using a 
phenomenological {\em pairing glue} given by
\begin{equation}
[c_{Z} \; + \; c_{\Delta} {\rm cos}(2(\phi \; - \; \phi^{'}))]F(\omega)
\end{equation}
This function replaces the standard $\alpha^{2}F(\omega)$ phonon spectral weight
in conventional Eliashberg theory. 
In applying the present model to high temperature superconductivity, 
$F(\omega)$ in equation (3) describes the frequency dependent spectral weight of
spin fluctuations that are assumed to be the origin of superconducting pairing. The $F(\omega)$ 
used for the present work are shown in figure (1)   

\begin{figure}
\begin{center}
\includegraphics[scale=1.0]{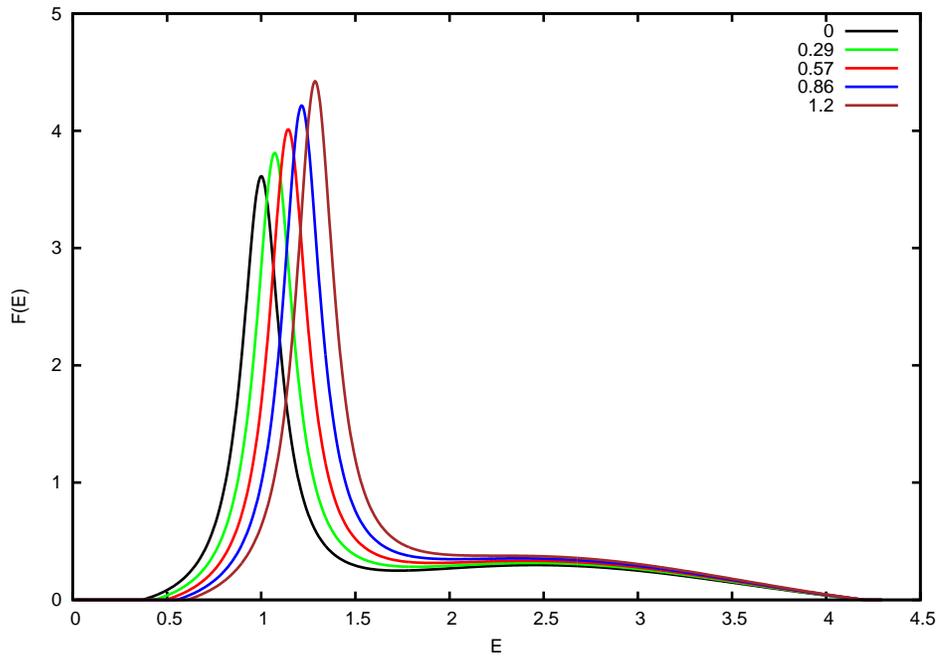}
\caption{The {\em pairing glue} spectral weight $F(\omega)$ for different
pseudogap values $\Delta_{0}^{PG}$ indicated on the figure.}
\label{}
\end{center}
\end{figure}

The prominent peak in $F(\omega)$ is the spin resonance mode measured in inelastic
neutron scattering (INS) which emerges on entering the superconducting state  in
Bi-2212 \cite{Brgs}.The position in energy of this peak in meV sets the energy scale, and magnitude 
in meV of quantities such as the superconducting gap, in the calculation.
The ${\rm cos}(2(\phi \; - \; \phi^{'})$ factor
leads to a d-wave symmetry superconducting gap described by
\begin{equation}
\Delta_{SC}(\nu,\phi) \; = \; \Delta(\nu) {\rm cos}(2 \phi)
\end{equation}
$\phi$ denotes the angular position on the Fermi surface. The function
$\Delta(\nu)$ is a complex function of frequency determined by the self
consistent solution of the equations (1) and (2).\\
\\
The $\phi$ dependent pseudogap $\Delta_{PG}(\phi)$ is modelled with an angle dependent function
which is a maximum at the antinode ($\phi = 0$), decreasing to $0$ at an angle $\phi_{C}$
before the d-wave node at $\phi= \pi/4$ is reached
\begin{equation}
\Delta_{PG}(\phi) \; = \; \Delta_{0}^{PG}(1 - (\phi/\phi_{C})^{2}) \; \; \; 0 \; < \; \phi \; < \; \phi_{C}
\end{equation}

The value of $\Delta_{0}^{PG}$ is a real valued input parameter.
$\phi_{C}$ is fixed at 0.45 radians.
Figure (2) depicts both the angular dependence of both gaps on the Fermi surface.

\begin{figure}
\begin{center}
\includegraphics[scale=1.0]{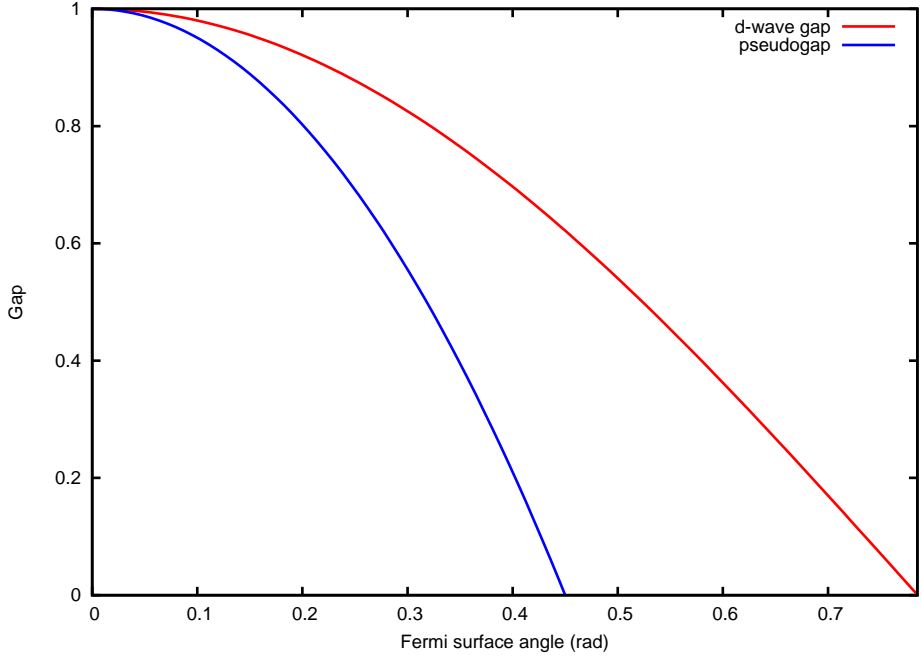}
\caption{Angle dependence ($\phi$) of the d-wave and pseudogap along the Fermi surface.}
\label{}
\end{center}
\end{figure} 

The resulting $\Delta_{SC}(\nu,\phi)$ are used to calculate the density of states
using
\begin{equation}
N(\nu) \; = \; \int_{0}^{2 \pi} \frac{d \phi}{2 \pi} 
{\rm Re} \bigg \{ \frac{\nu}{[ \nu^{2} - \Delta^{2}(\nu){\rm cos}^{2}(2 \phi) - \Delta_{PG}^{2}(\phi) ]^{1/2}} \bigg \}
\end{equation}
The SIS current-voltage curve is calculated the usual way, using a tunnelling 
density of states $N^{T}(\nu)$ that includes a directional tunneling matrix \cite{LC}
\begin{equation}
I(eV) \; = \; \int^{\infty}_{-\infty} d \nu N^{T}(\nu +eV)N^{T}(\nu)[f( \nu )-
f( \nu +eV)]
\end{equation}
The role of the tunneling directionality factor is required to generate the large peak height
to high bias background that is actually seen in the SIS break junction experiments.
The SIS conductance curves (dI/dV) are obtained by numerical differentiation.\\
\newpage
{\bf Results}\\
\\
Figures (3) and (4) show weak coupling calculations of
the density of states, and the corresponding SIS conductance (dI/dV), for
the case of a combined d-wave superconducting gap and pseudogap.
The superconducting gap function $\Delta(\nu)$ is set equal to a real constant $\Delta_{0}$ in equations (6) and (7). The values chosen for 
$\Delta_{0}$ and the pseudogap parameter $\Delta _{0}^{PG}$ (see equation (5))
are indicated on the figures. \\
\\
For $\Delta_{0} > \Delta_{0}^{PG}$, the presence of a non-zero pseudogap results in a small break in
slope at an energy below the main peaks in both figure (3) and (4). Therefore, for this case, 
there is a small impact due to the presence of a pseudogap on the
shape of the density of states, or the SIS conductance.
When $\Delta_{0}  <  \Delta_{0}^{PG}$, a {\em shoulder}
feature is visible in the density of states at an energy below that of the main peak in figure (3).
This is the first noticeable evidence in the density of states 
indicating the presence of two gaps. The
location of the main peaks in both density of states and SIS conductance in figures (3) and (4) are determined by the factor $( (\Delta_{0})^{2} + (\Delta_{0}^{PG})^{2})^{1/2}$. Therefore,
identifying the energy of either the superconducting gap, or the pseudogap, with the energy of
the peak is correct only in the special cases of zero superconducting gap or
zero pseudogap. When both gaps co-exist, the peak is determined by a combination of both gap
values. This would be true not only for the present model for incorporating the pseudogap but
also in other models \cite{Tmr}. 
Note that
the {\em shoulder} that develops below the main SIS peak in figure (4) is unrelated to a 
similar {\em shoulder}
feature in the density of states seen in figure (3). 
This can be seen clearly in figure (5) where the conductance (dI/dV) from equation (7) is shown
for a pure pseudogap case (with $\Delta_{0}^{PG} = 1.7$). The energy of the {\em shoulder}
feature in figure (5) is at half the energy of the main peak (located at $2\Delta_{0}^{PG} $). 
The {\em shoulder}
feature in figures (4) and (5) arises from the angular dependence of the pseudogap (see equation (5)). 
Similarly, there is also a small
break in the slope of the SIS dI/dV for a pure superconducting d-wave case at the energy of half the main SIS peak.
Weak coupling calculations\cite{Carb2} of the density of states using the electron/hole pocket
model \cite{Tmr} display similar {\em shoulder/small peak} features. \\
\\
\begin{figure}
\begin{center}
 \includegraphics[scale=1.0]{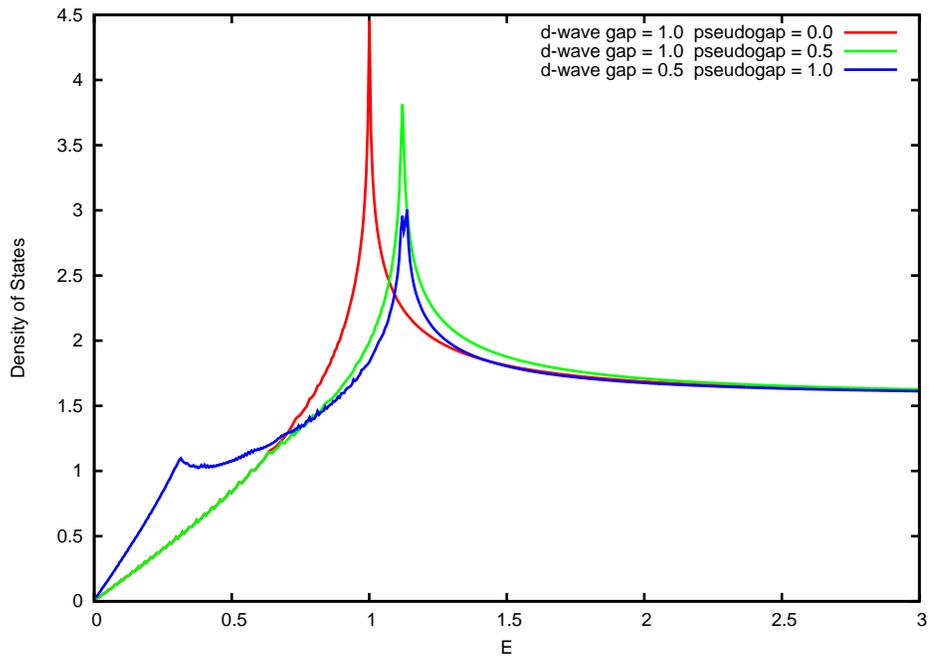}
 \caption{Weak coupling density of states}
  \label{}
 \end{center}
 \end{figure}

 \begin{figure}
\begin{center}
 \includegraphics[scale=1.0]{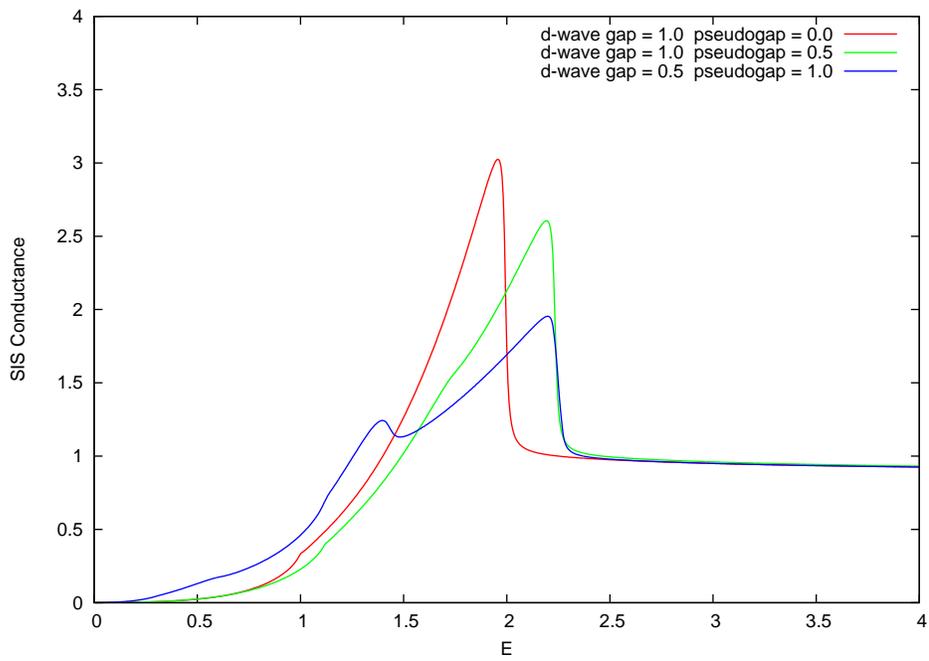}
 \caption{Weak coupling SIS conductance dI/dV }
 \label{}
 \end{center}
 \end{figure}
 
\begin{figure}
\begin{center}
\includegraphics[scale=1.0]{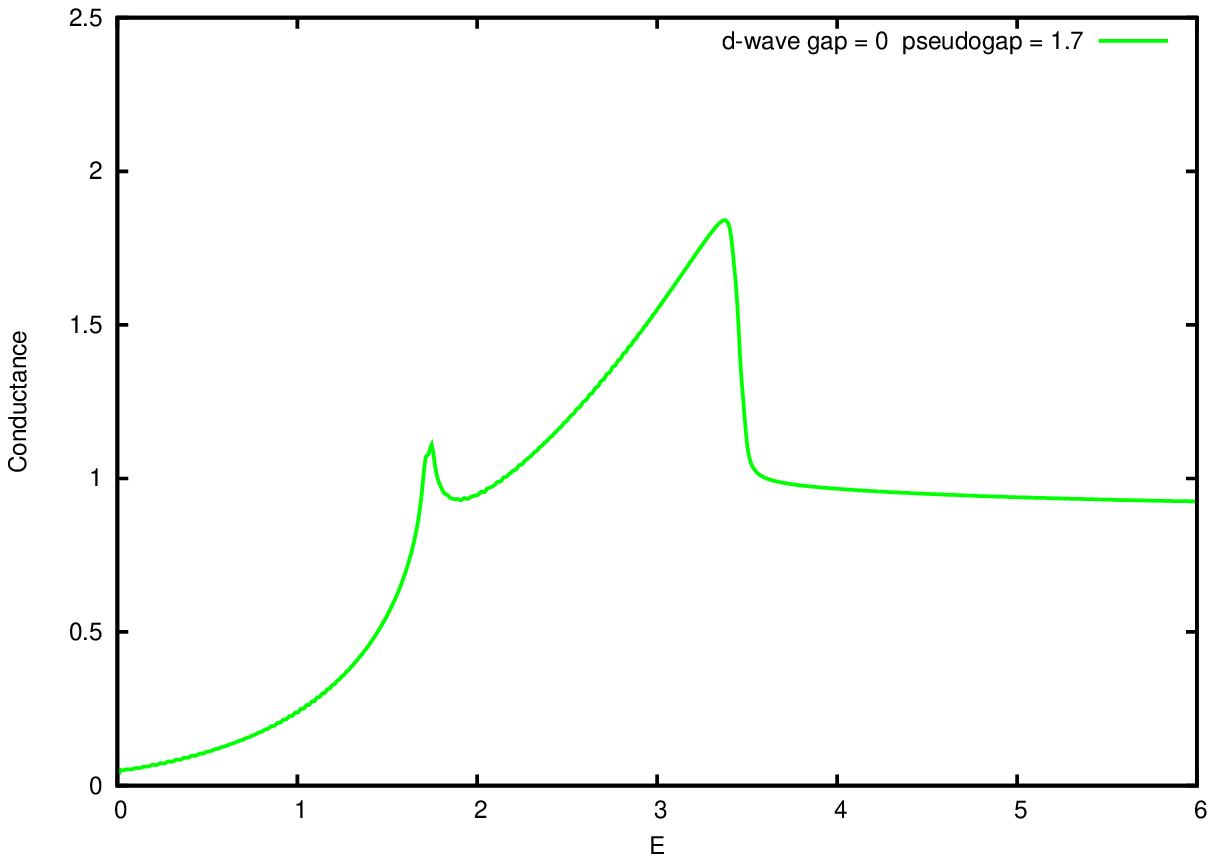}
\caption{Conductance dI/dV for pure pseudogap state using eqn (7) }
\end{center}
\end{figure}

Strong coupling results are shown in figures (6), (7) and (8) for the superconducting gap, the density of states, and
the SIS conductance using equations (1) and (2) for a range of values of the pseudogap parameter $\Delta_{0}^{PG}$.\\
\\
In the present work, the pseudogap $\Delta_{0}^{PG}$ is assumed to start at zero at optimal doping
in the case of Bi-2212, and to increase monotonically upon moving into the underdoped regime. 
Figure (6) shows the effect of the increasing pseudogap on the superconducting gap. 
The superconducting gap being referred to here is obtained from the usual
strong coupling criterion as the energy $E$ where $E= {\rm Real} \Delta(E)$.\\

The starting case $\Delta_{0}^{PG}=0$, along with the $F(\omega)$ for $\Delta_{0}^{PG}=0$ shown
in Figure (1), was used previously to describe optimal doped Bi-2212 \cite{LC}.
In this case, the position of
the peak in the $F(\omega)$ of figure (1) corresponds to 40meV which is the value approximately
of the spin resonance mode energy detected by INS in Bi-2212 \cite{Brgs}. The shape of $F(\omega)$, and the values of
$c_{S}=0.14$ and $c_{\Delta}=0.8$, yield a superconducting gap of slightly less than 40meV for this optimally
doped case in agreement with SIS measurements \cite{JZ}.\\
\\
The superconducting gap is suppressed to zero when the pseudogap reaches 1.7 in the units
shown in figure (6). Again, applying this to Bi-2212, where 40 meV corresponds to unity in figure (6), this would
mean a pseudogap of about 70meV coincides with the suppression of $T_{C}$, and the corresponding 
superconducting gap, to zero, consistent with experimental results\cite{Huf}.  

\begin{figure}
\begin{center}
\includegraphics[scale=1.0]{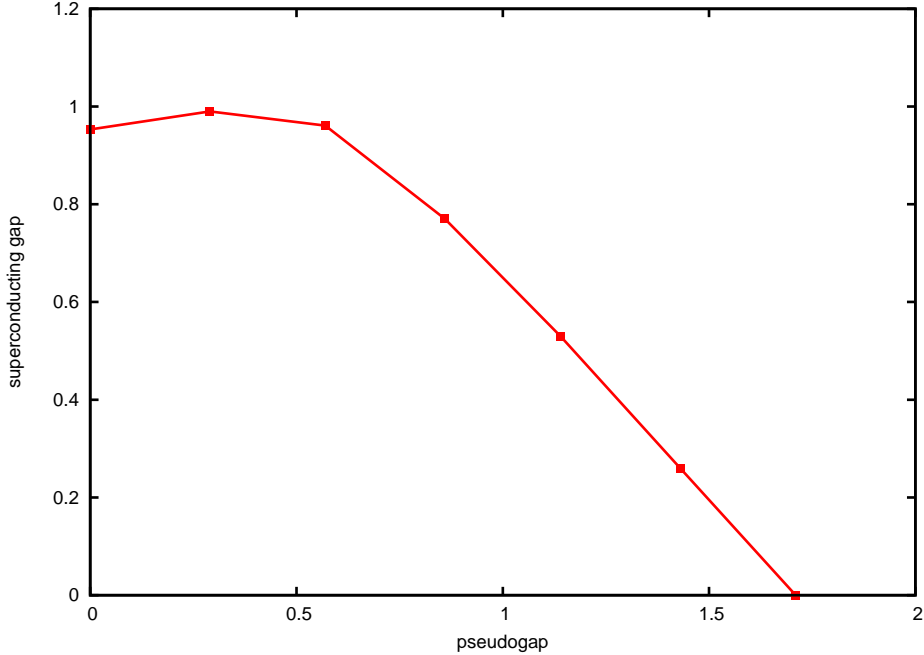}
\caption{Superconducting gap versus increasing pseudogap $\Delta_{0}^{PG}$ (eqn (5))}
\label{}
\end{center}
\end{figure}

The results for increasing pseudogap $\Delta_{0}^{PG}$ shown in figures (6), (7) and (8) are calculated using the
$F(\omega)$ of figure (1). The coupling constants are
kept at the constant values $c_{S}=0.14$ and $c_{\Delta}=0.8$ throughout the calculations
The upward shift in energy of the main peak in
$F(\omega)$ spectral weight, as shown in figure (1) is similar to
theoretical results for the spin fluctuation Im$\chi(\omega)$ in numerical studies on the
Hubbard model \cite{Trem}. (see figure (1)(b) and accompanying discussion in reference (18)).
\\
Figure (7) shows the evolution of the density of states as the superconductor evolves from the optimally
doped case ($\Delta_{0}^{PG}=0$) to complete suppression of superconductivity ($\Delta_{0}^{PG}=1.7\simeq 70$meV (Bi-2212)).
As the superconductivity is suppressed, the peak in the density of states moves to higher energy, driven by the
increasing pseudogap energy. A {\em shoulder} feature develops at low energy, the position of which is related to
the magnitude of the self consistently generated superconducting gap. The position of this feature is not at the
same energy as the superconducting gap, however. The dip feature gradually weakens, and disappears in the pure
pseudogap state. These trends are similar to the STM experiments.\\
\\
Figure (8) shows the SIS conductance curves corresponding to figure (7). The main SIS peak increases with energy, driven
by the increasing pseudogap, and also broadens significantly. The dip feature above the main peak weakens, and is
not present in the pure pseudogap conductance. This latter behavior is seen in SIS experiments \cite{JZ} in the
large pseudogap (very underdoped) state. This indicates
that the physics of the pseudogap formation is not coupled to the physical mechanism causing the dip feature in
the superconducting state, which in the latter case is the spin resonance mode. The absence of the dip feature in
the SIS experimental measurements also lends support to the present model where a frequency independent pseudogap, defined in equation (5), is used in the model. The {\em shoulder} feature that is present in the curves in figure(8) is not seen in the experimental SIS peaks \cite{JZ} possibly due to
the significant broadening of the experimental peaks. \\
\\ 
 
\begin{figure}
\begin{center}
\includegraphics[scale=1.0]{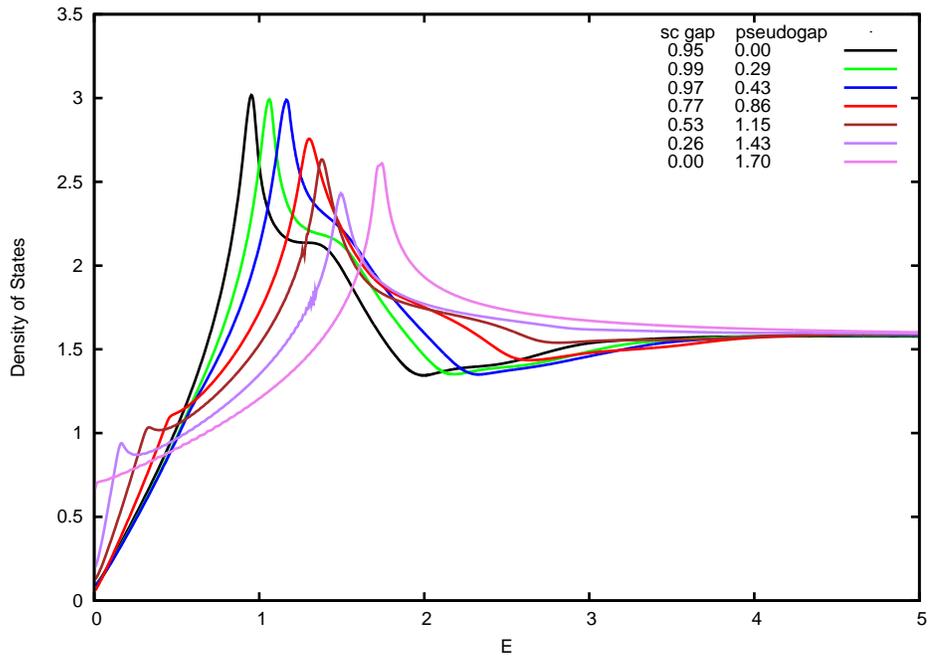}
\caption{Strong coupling density of states (eqn (6))}
\label{}
\end{center}
\end{figure}

\begin{figure}
\begin{center}
\includegraphics[scale=1.0]{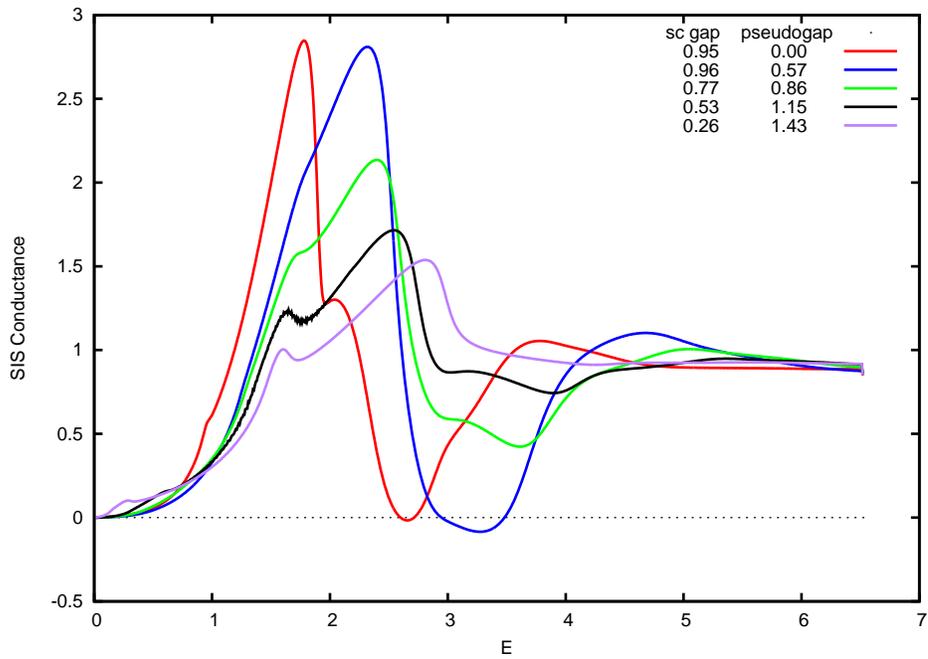}
\caption{Conductance dI/dV (eqn (7)) corresponding to Figure (7)}
\end{center}
\end{figure}

\newpage

Figure (9) shows a plot for comparison of the superconducting gap magnitude calculated from the strong coupling equations (see figure (6)) and the gap magnitude obtained from taking the energy of the main peak in the SIS conductance curves (figure (8)) and dividing by two. The two curves diverge from each other as the pseudogap increasingly determines the peak position in figure (8) for larger
pseudogap magnitudes (deeper into the underdoped region).\\
\\

{\bf Summary}\\
\\
A strong coupling model, which has been used previously to analyze
SIS break junction experiments on optimal to overdoped Bi-2212 \cite{LC}, is modified to include a
pseudogap in the electronic spectrum. The resulting density of states and
SIS conductance curves compare favorably with experimental measurements on
underdoped Bi-2212. Underdoping is simulated in the calculation by
increasing the magnitude of the input pseudogap, assuming that the
pseudogap increases as doping is decreased in the underdoped state.

\begin{figure}
\begin{center}
\includegraphics[scale=1.0]{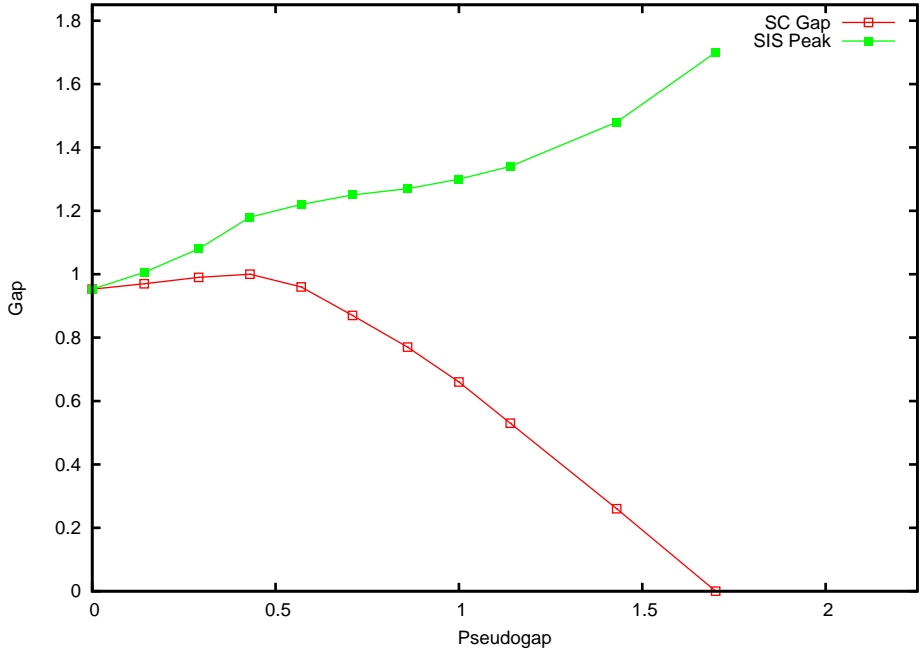}
\caption{Comparison of strong coupling superconducting gap (red curve) (plotted in Figure (6)) and half the energy of the SIS peak from Figure (8) (green curve)}
\end{center}
\end{figure}

\end{document}